\documentclass[%
 aip,
 jmp,%
 amsmath,amssymb,
 reprint,%
]{revtex4-1}

\usepackage{graphicx}
\usepackage{dcolumn}
\usepackage{bm}
\usepackage{hyperref}
\begin{document}

\title{Anomalous thermal expansion of Sb$_2$Te$_3$ topological insulator}

\author{P. Dutta}
\author{D. Bhoi}
\author{A. Midya}
\author{N. Khan}
\author{P. Mandal}
\address{Saha Institute of Nuclear Physics, 1/AF Bidhannagar, Calcutta 700 064, India}

\author{S. Shanmukharao Samatham}
\author{V. Ganesan}
\address{UGC-DAE Consortium for Scientific Research,
Khandwa Road, Indore 452 001, India}
\date{\today}
\begin{abstract}
We have investigated the temperature dependence of the linear thermal expansion along the hexagonal $c$ axis ($\Delta L$), in-plane resistivity ($\rho$), and specific heat ($C_p$) of the topological insulator Sb$_2$Te$_3$ single crystal. $\Delta L$ exhibits a clear anomaly in the temperature region 204-236 K. The coefficient of linear thermal expansion $\alpha$ decreases rapidly above 204 K, passes through a deep minimum at around 225 K and then increases abruptly in the region 225-236 K. $\alpha$ is negative in the interval 221-228 K. The temperature dependence of both $\alpha$ and $C_p$ can be described well by the Debye model from 2 to 290 K, excluding the region around the anomaly in $\alpha$.
\end{abstract}
\pacs{}
\maketitle
\newpage

Topological Insulators (TIs), being a topic with fascinating physics and high application potential, have been a highly emerging field in last five years \cite{hasan}. These materials possess an insulating energy gap in the bulk due to the spin-orbit coupling and gapless surface states protected by the time-reversal symmetry. The topologically protected surface states host a variety of properties like robust conductance, relativistic Dirac cone, spin-momentum chirality, and quantized anomalous Hall effect \cite{hasan,qi}. These dissipationless surface states are potentially useful for applications in spintronics and quantum computations, and also in exploring the fundamental physics like Majorana Fermions, magnetic monopoles and axion electrodynamics. Though, many members of TIs such as Bi$_2$Se$_3$, Bi$_2$Te$_3$, Sb$_2$Te$_3$, etc. have been studied earlier as good thermoelectric materials, it is essential to characterize their basic physical properties more accurately down to the lowest temperature for device designing. The atomic arrangement of TIs is a layered structure, with each layer consisting of five monoatomic planes of Te(Se)-Sb(Bi)-Te(Se)-Sb(Bi)-Te(Se). Each of these five atomic planes is referred to as a quintuple layer (QL). The lattice properties of these materials are of considerable interest because of the weak van der Waals interaction between the adjacent QLs. Previous reports  show that the thermal expansion in these materials is quite anisotropic  and exhibits an anomalous temperature dependence, presumably because of the weak bonding \cite{barnes,fran,tay,chen}. Chen \emph{et al.} \cite{chen} studied the temperature dependence of the lattice parameters of Bi$_2$Se$_3$ and Sb$_2$Te$_3$ single crystals. Below 150 K, the  extracted coefficient of linear thermal expansion $\alpha$ can be described well by the Debye model. While above 150 K, the estimated value of $\alpha$  is larger than the theoretical one and does not show saturation-like behavior well above the Debye temperature. Furthermore, from the Raman Spectroscopy measurements, it has been shown that the thermal expansion contribution term $\Delta\omega(T)$ accounted for 40\% of the total phonon frequency change with temperature \cite{kim}.\\

In most cases, as the $\alpha$ was determined from the temperature variation of the lattice parameters, the magnitude and nature of the $T$ dependence of the reported anomalies differ significantly from each other due to the lack of accuracy. Thus an accurate measurement of thermal expansion is necessary to shed some more light on the nature of anomaly and see whether it has any correlation with electronic properties. In this regard, we have measured the temperature dependence of the in-plane resistivity ($\rho$), specific heat ($C_p$) and linear thermal expansion ($\Delta L$) along $c$ axis of Sb$_2$Te$_3$ single crystal. We observe that $\Delta L$($T$)  exhibits an anomaly in the region 204-236 K. No anomaly is observed in $C_p$($T$) and $\rho$($T$) data. Both $C_p$($T$) and  $\alpha$($T$) can be described well by the Debye model.\\

\begin{figure}[b]
  \includegraphics[width=0.45\textwidth]{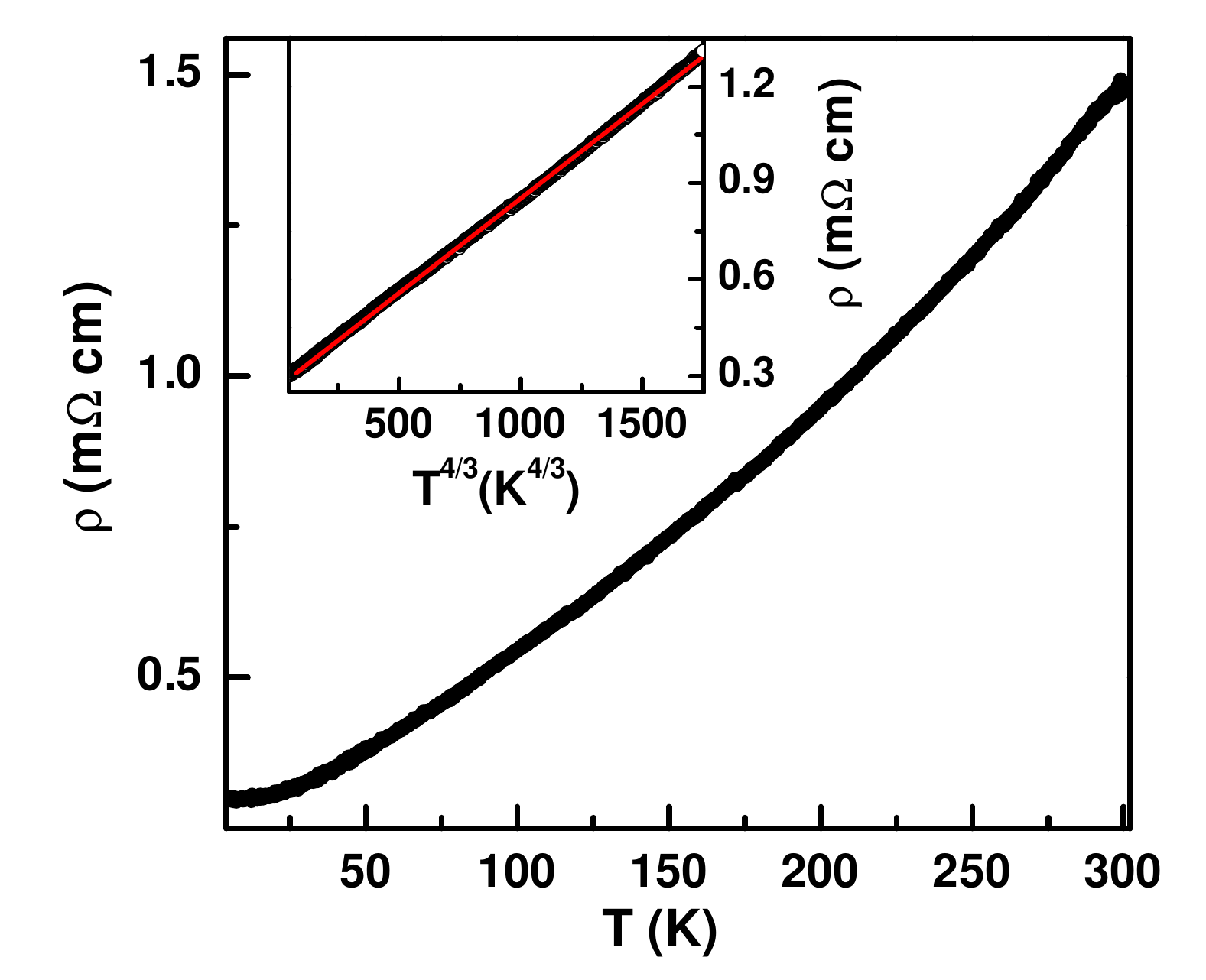}
  \caption{Temperature dependence of the in-plane  resistivity $\rho$ of the Sb$_2$Te$_3$ single crystal. Inset shows the power-law fit, $\rho(T)=\rho_0 + aT^n$ with $n$$\sim$4/3, to the resistivity data in the temperature region 20-270 K.}\label{Fig1}
\end{figure}
Single crystals of Sb$_2$Te$_3$ were grown by melting the stoichiometric mixture of elemental Sb (99.999\%) and Te (99.999\%) at 850 $^{\circ}$C for 24 h in a sealed vacuum quartz tube. The sample was cooled over a period of 48 h to 620 $^{\circ}$C and kept at this temperature for 96 h before quenching in liquid nitrogen. Crystals of millimeter size with shiny flat surface can easily be obtained from the boule of Sb$_2$Te$_3$. The phase purity of the sample was checked by powder x-ray diffraction (XRD) method with CuK$_{\alpha}$ radiation in a Rigaku x-ray diffractometer (TTRAX II) and no impurity phases were detected. The XRD pattern can be fitted well with a hexagonal structure with space group $R\bar{3}m$. The evaluated lattice parameters are $a_{hex}$ = 4.2654 $\AA$ and $c_{hex}$ = 30.4438 $\AA$. Four probe measurements of the electrical resistivity between 2 and 300 K were done using a commercial cryostat (Cryogenic Ltd.). For the thermal expansion measurement, a miniature tilted plate capacitance dilatometer was utilized which allows an accurate study of sample length change \cite{rotter}. We have measured the macroscopic length change $\Delta L(T)$ of a crystal of  $\sim$1 mm thickness and calculated the coefficient of linear thermal expansion,
\begin{equation}
\alpha = \frac{1}{L_0}.\frac{\partial(\Delta L)}{\partial T},
\end{equation}
where $L_0$ is the length of the sample at lowest measured temperature. The specific heat  was measured by the relaxation time method using a Quantum Design physical property measurement system. \\

Figure 1 shows the temperature dependence of  $\rho$ for Sb$_2$Te$_3$ single crystal. With decreasing $T$, $\rho$ decreases continuously down to 20 K and then saturates to a constant value. No anomaly has been observed in the measured temperature range. Both the magnitude and $T$ dependence of $\rho$ for the present sample are comparable to that for Bi$_2$Se$_3$ and Bi$_2$Te$_3$ crystals with carrier density $10^{18}$-$10^{19}$ cm$^{-3}$ \cite{butch,ando,james}. In Bi$_2$Se$_3$, for carrier density greater than $10^{18}$ cm$^{-3}$, $\rho(T)$ data reflect good metallic behavior  \cite{butch}. In comparison to the above mentioned crystals, the residual resistivity ratio $\rho_{300 \rm{K}}/\rho_{2 \rm{K}}\sim$5 for the Sb$_2$Te$_3$ crystal is higher \cite{butch,ando,james}. In the temperature region 20-270 K, the resistivity  can be fitted well by a power-law expression, $\rho(T)$=$\rho_0 + aT^n$ with exponent $n$$\sim 4/3$ as shown in the inset of Fig. 1. The observed value of the exponent does not seem to classify this compound into hither-to-known metallic systems. Normally, in metallic elements, alloys, and compounds, $\rho$ shows $T^2$ dependence at low temperature and $T$-linear behavior at high temperature. In magnetic materials, $\rho$ may show $T^{3/2}$ dependence at low temperature due to the magnetic contribution to the scattering. In the bulk samples of TIs, the transport properties of the surface states are often mixed with those from the bulk states making it difficult to observe insulating behavior in $\rho(T)$. As the bulk conductivity  decreases whereas the conductivity due to the surface states increases with decreasing temperature, the electrical conduction at low temperature is dominated by the surface.\\
\begin{figure}
  \includegraphics[width=0.45\textwidth]{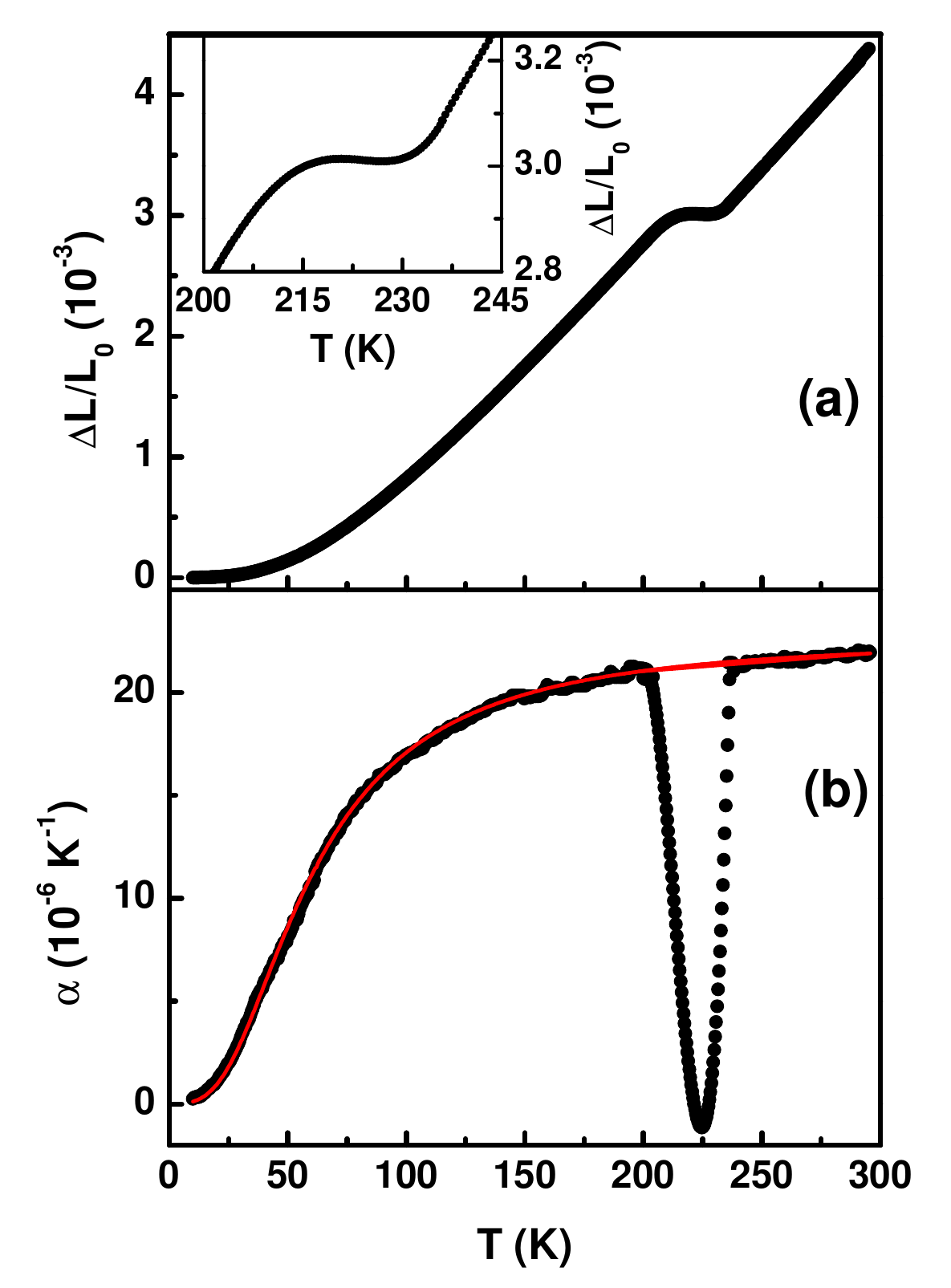}
  \caption{(a) Temperature dependence of the linear thermal expansion $\Delta$$L/L_0$ along the $c$ axis of the Sb$_2$Te$_3$ single crystal. Inset shows the expanded view of the anomalous region. (b) Temperature dependence of the  coefficient of linear thermal expansion $\alpha$. The solid line is the fit to the Debye model [Eqn.(2)].}\label{Fig2}
\end{figure}

Figure 2(a) shows the temperature dependence of $\Delta L/L_0$ along the hexagonal $c$ axis of Sb$_2$Te$_3$ crystal in the range 4 to 290 K. We have also measured the thermal expansion in presence of magnetic field but $\Delta L/L_0$ does not show any $H$ dependence up to 7 T. The length decreases with decrease in $T$ as it occurs in most metals, however, $\Delta L/L_0$ exhibits a broad minimum at $\sim$228 K followed by a broad maximum at $\sim$221 K [inset of Fig. 2(a)]. With further decrease in $T$, $\Delta L/L_0$ decreases continuously down to the lowest measured temperature. This indicates that there exists a small temperature region 221-228 K where an anomalous thermal expansion occurs. The anomalous behavior is clearly reflected in the coefficient of linear thermal expansion. In Fig. 2(b), $\alpha$ is shown as a function of $T$. Except for 204 K$\leq$$T$$\leq$236 K, $\alpha$ increases smoothly with the increase of $T$ and saturates at high temperature. Above 204 K, $\alpha$ decreases sharply with increasing $T$, passes through a deep minimum at around 225 K and then increases abruptly in the region 225-236 K. For 221 K$\leq$$T$$\leq$228 K, $\alpha$ is negative. Though the anomaly in the region 204-236 K is slightly asymmetric, it is not a $\lambda$-like as in the case of a  second-order phase transition. Excluding the region around the anomaly, the $\alpha(T)$ curve can be fitted well in the whole temperature region with the Debye model,
\begin{equation}
\alpha(T)= \alpha_0\left(\frac{T}{\Theta_D}\right)^3\int_0^{\Theta_D/T}\frac{x^4 e^{x}dx}{(e^{x}-1)^2},
\end{equation}
where $\alpha_0$ is $T$-independent fitting parameter and $\Theta_D$ is the Debye temperature. The estimated value of $\Theta_D$ is 235 K. Chen \emph{et al.} \cite{chen} determined $\alpha$ from the temperature dependence of lattice parameters of Sb$_2$Te$_3$. They observed that $\alpha$($T$) can be described well by the Debye model up to 150 K with $\Theta_D$$\sim$200 K. However, above 150 K, $\alpha$ along $c$ axis initially decreases slowly with $T$ down to 180 K and then increases, as a result, the experimental value of $\alpha$ is higher than the Debye model for $T$ $>$ 230 K. On the contrary, $\alpha$  perpendicular to the $c$ axis increases monotonically with $T$. This suggests that the large anomaly in thermal expansion along the $c$ axis is associated with the inter-layer van der Waals interaction.\\
\begin{figure}
  \includegraphics[width=0.45\textwidth]{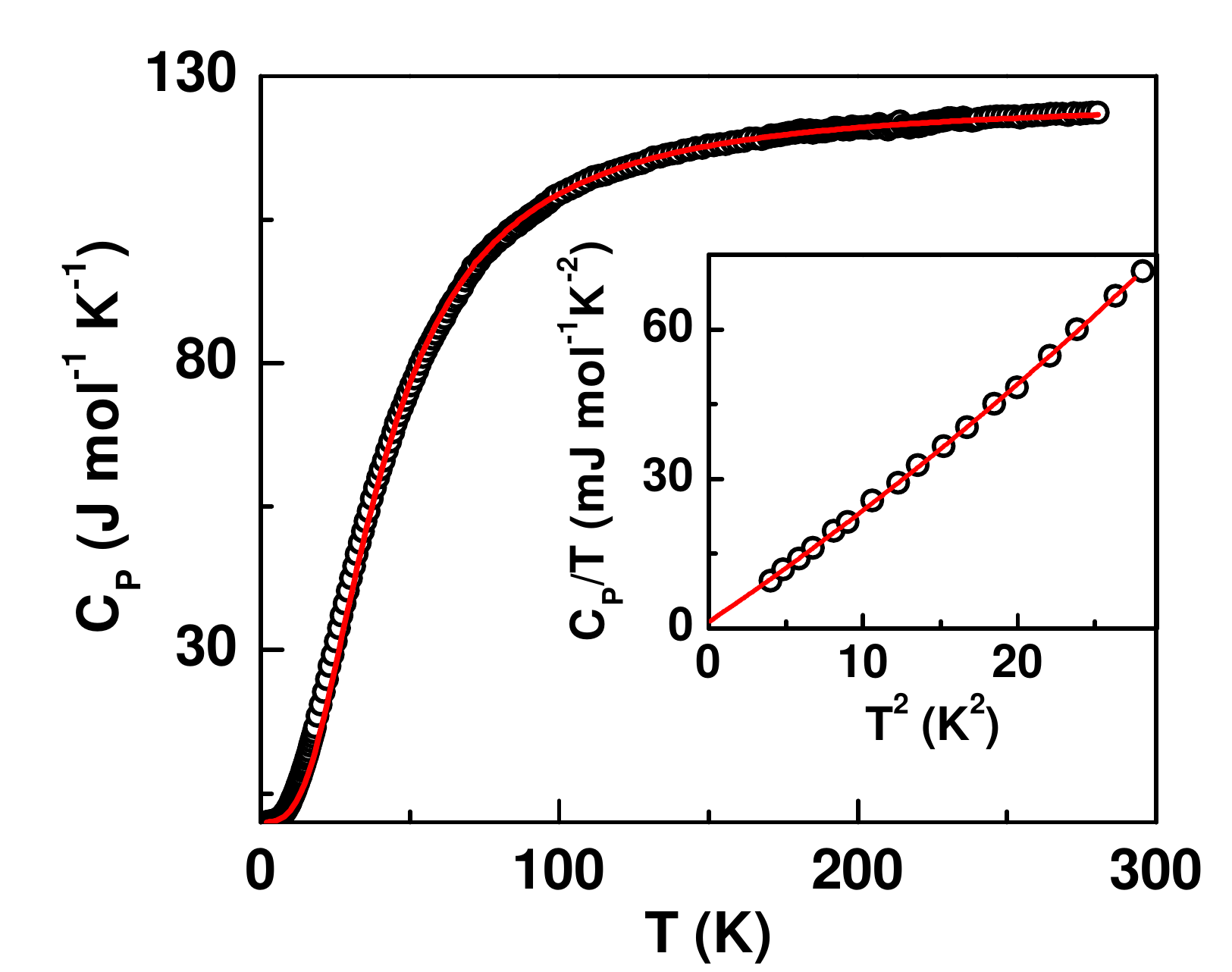}
  \caption{Temperature dependence of the specific heat of the Sb$_2$Te$_3$ single crystal. The solid line is the fit to the Debye model (see the text). Inset: $C_p/T$ data below 5 K has been fitted (solid line) with the expression $C_p/T=\gamma  + A_3 T^2 + A_5 T^4$.}\label{fig3}
\end{figure}

In order to investigate whether this anomaly in thermal expansion is due to the phase transition, we have measured the specific heat of Sb$_2$Te$_3$ from 2 to 300 K as shown in Fig. 3. The $C_p$($T$) data  do not show any anomaly around the temperature interval where the anomalous thermal expansion is observed, ruling out the possibility of  any electronic or structural phase transition. Also, one can see that $C_p$ tends to saturate at $\sim$125 J mol$^{-1}$ K$^{-1}$ at high temperature which is close to the Dulong-Petit value. Both the value and temperature dependence of  $C_p$  for the present sample match very well with the reported specific heat data for Sb$_2$Te$_3$ \cite{pashin}. Pashinkin \emph{et al.} \cite{pashin} measured $C_p$ on Sb$_2$Te$_3$ single crystal over a wide range of $T$  above 350 K and observed  a very weak $T$ dependence in $C_p$. The specific heat data in Fig. 3 can be described well by the Debye model (Eqn. 2) where $\alpha_0$ is replaced by 9$NR$ and an extra term $\gamma$$T$ due to electronic contribution is added, where $\gamma$ is the Sommerfeld coefficient. Also, the estimated value of $\Theta_D\sim$165 K agrees very well with that reported in literature \cite{madelung}. For the accurate determination of $\gamma$, the specific heat data below 5 K can be described by, $C_p = C_{el} + C_{ph}$ = $\gamma T + A_3 T^3 + A_5 T^5$, where $A_3$ and $A_5$ are the coefficients of the phononic contribution (inset of Fig. 3) \cite{shoe}.  From the fit, the  value of $\gamma$ can be confined to 0.7$\pm$0.7 mJ mol$^{-1}$ K$^{-2}$.\\

In conclusion, we have investigated the temperature dependence of the resistivity, specific heat, and linear thermal expansion of the topological insulator Sb$_2$Te$_3$. $\Delta$$L$ exhibits an anomalous thermal expansion in the temperature region 221-228 K. $\rho(T)$ shows an unusual power-law behavior for 20 K$\leq$$T$$\leq$270 K. Both $\alpha$($T$) and $C_p$($T$) can be described well by the Debye model excluding the region where anomaly is observed in $\alpha$. \\

We would like to thank A. Pal and S. Das for their help during the sample preparation and measurements. P. Dutta would  like to thank CSIR, India for Junior Research fellowship (File no. 09/489(0086)/2010-EMR-I). V. Ganesan also thanks Director and Centre Director, UGC-DAE CSR, Indore for their support and the DST for the LTHM project.

\end{document}